\RequirePackage{fix-cm}
\documentclass[twocolumn]{svjour3}          
\smartqed  
\usepackage{graphicx}
\usepackage{hyperref}
\usepackage{amsfonts}
\usepackage{amssymb}
\usepackage{amsmath}
\usepackage{graphicx}
\usepackage{epsfig}
\usepackage{color}
\journalname{Brazilian Journal of Physics}
\begin{document}

\title{Eliminating the cuspidal temperature profile of a non-equilibrium chain \thanks{We would like to thank the partial funding from FINEP [contract No. PUC-Infra 1580/10], FAPERJ [contract No. APQ1-110.635/2014] and CNPq [contract No.s 481640/2011-8
and 308737/2013-0].
This work benefited from financial support from and FAPERJ
[Contracts {\em Apoio a Projetos Tem\'{a}ticos no 
Estado do Rio de Janeiro}  210.958/2015 and {\em Jovem Cientista do nosso 
Estado} 202.881/2015] and CNPq [Contract No. 308737/2013-0] (SMDQ). W. A. M. M. acknowledges CNPq (481640/2011-8) and FAPERJ (APQ1 - E-26/111.455/2008) for partial financial support. M. M. C acknowledges CNPq for financial support. 
}
}

\titlerunning{Eliminating the cuspidal temperature profile }       

\author{Michael~M. C\^{a}ndido       \and
        Welles A.~M. Morgado \and 
        S\'{\i}lvio M. Duarte~Queir\'{o}s.
}

\institute{M.~M. C\^{a}ndido  \at
              Department of Physics, PUC-Rio \\
            Rua Marqu\^es de S\~ao Vicente 225, 22453-900 Rio de Janeiro\\
           \and
           W.~A.~M. Morgado  \at
              Department of Physics, PUC-Rio and \\
            National Institute of Science and Technology for Complex Systems \\
            Rua Marqu\^es de S\~ao Vicente 225, 22453-900 Rio de Janeiro\\
              \email{welles@puc-rio.br}           \\
           \and
           S.~M. Duarte~Queir\'{o}s  \at
             Centro Brasileiro de Pesquisas F\'{\i}sicas and\\
            National Institute of Science and Technology for Complex Systems \\
            Rua Dr Xavier Sigaud, 150, 22290-180 Rio de Janeiro --- RJ, Brazil \\
}

\date{Received: date / Accepted: date}

\maketitle

\begin{abstract}
In 1967, Z.~Rieder, J.~L. Lebowitz and E.~Lieb (RLL) introduced a model of heat conduction on a crystal that became a milestone problem of non-equilibrium statistical mechanics. Along with its inability to reproduce Fourier's Law --- which subsequent generalizations have been trying to amend --- the RLL model is also characterized by awkward cusps at the ends of the non-equilibrium chain, an effect that has endured all these years without a satisfactory answer.
\\
In this paper, we first show that such trait stems from the insufficiency of pinning interactions between the chain and the substrate. Assuming the possibility of pinning the chain, the analysis of the temperature profile in the space of parameters reveals that for a proper combination of the border and bulk pinning values, the temperature profile may shift twice between the RLL cuspidal behavior and the expected monotonic local temperature evolution along the system, as a function of the pinning. At those inversions, the temperature profile along the chain is characterized by perfect plateaux:
at the first threshold, the cumulants of the heat flux reach their maxima and the vanishing of the two-point velocity correlation function for all sites of the chain so that the system behaves similarly to a ``phonon box''. On the other hand, at the second change of the temperature profile, we still have the vanishing of the two-point correlation function but only for the bulk, which explains the emergence of the temperature plateau and thwarts the reaching of the maximal values of the cumulants of the heat flux.
\end{abstract}

\section{Introduction}
\label{intro}

The problem of heat conduction is certainly one of the best examples for illustrating the role of statistical mechanics in the treatment of non-equilibrium systems~\cite{nakazawa-70,lepri-03,dhar-08,bonetto-00}. It is a manifestation of the laws of thermodynamics, well summarised  by Fourier's law: the heat flux density is equal to minus the product of the thermal conductivity by the temperature gradient~\cite{kubo-91},
\begin{equation}
\vec{J} = - \kappa \, \nabla T \,.
\label{fourier-law}
\end{equation}

Since matter is discontinuous, the constituent particles of a system interact in a way that has obvious implications on the macroscopic effects we observe. Actually, as early as the introduction of the kinetic theory by Boltzmann, there have been several attempts to microscopically derive and explain the emergence of Fourier's law and, thereafter, the same endeavor was carried out within the context of condensed matter physics~\cite{ashcroft-76,chaikin-95}.

Although the actual microscopic understanding of a physical system requires the application of a quantum mechanical approach, several classical (toy) models have been introduced --- especially from the 1950s on~\cite{bergmann-55,lebowitz-67,nakazawa-70} --- aiming at studying the problem of heat transport in crystals, via the assumption of $d$-di\-men\-si\-o\-nal lattices of coupled oscillators in contact with reservoirs at different temperatures, $T_C$ and $T_H > T_C$, placed at each end of the chain. For $d=1$, the most emblematic  of the models is the one introduced by  Rieder, Lebowitz and Lieb (RLL)~\cite{lebowitz-67}. Unsurprisingly at all, this model reaches a steady state without abiding by Fourier's law, \emph{i.e.}, due to the absence of phonon scattering, it has an infinite heat conductivity corresponding to ballistic heat transport~\cite{lebowitz-67}.  A finite heat conductivity for the homogeneous case (identical particles and springs)  can be obtained for higher dimensional non-linear systems~\cite{saito-10}. For the one-dimensional case, several modifications have been introduced~\cite{kannan-13,mimnagh-97,linear}, from considering mass dispersion~\cite{huang-06,yang-07}, free-particles and pinned particles (ding-a-ling)~\cite{casati-84}, extra reservoirs along the chain~\cite{lepri-98,jani-04} to effective external collisions~\cite{olla,dhar-11,landi-13}, among a myriad of other variants.

At the same time, the RLL proposal is characterized by a temperature profile with a cusp-anticusp on the edges of the chain. Notwithstanding all the effort put into bringing forth different ways of giving rise to Fourier's law, when we look at the temperature profiles of the RLL variants, we verify that the bump (dip) in the vicinity of the colder(hotter) reservoir is still present. Intuitively, that behavior is quite odd and prom\-pt\-ed the authors of Ref.~\cite{lebowitz-67} to mention it on the abstract; let us notice that such result simply means the warmer (colder) particles are closer to the colder (hotter) reservoir whereas common sense tells us that the temperature profile should change monotonically across the chain, even when Fourier's Law is not verified --- but heat is still  transported superdiffusively --- and allowing the determination of a bulk temperature $\mathcal{T} \equiv \left( T_C+T_H \right) / 2$.

The (starting) aim of this manuscript is therefore to provide answers to the following questions that have endured for 50 years: 

\hspace{-0.75cm}
1) \emph{Why does the Rieder-Lebowitz-Lieb model of heat conduction exhibit a cuspidal temperature profile? How can we change that oddity and obtain a smooth temperature profile?} 

\hspace{-0.75cm}
In the wake of answering these two intermingling questions we go further afield and discuss the following issue: 

\hspace{-0.75cm}
2) \emph{What does happen to the thermostatistical features of the mechanical system as we change it to smooth the temperature profile?}

The remaining of our manuscript is organized as follows: in the next section, Sec.~\ref{model}, we will introduce the model describing the system and the method we will employ to obtain the solutions;  in Sec.~\ref{results}, we will answer the two questions we have asked in the last paragraph; and in Sec.~\ref{remarks}, we will give an overall picture of our results and their impact on the treatment of problems of heat conduction.

\section{Model and method of solution}
\label{model}

Our study revolves around the dynamics of a chain of $N$ linearly coupled oscillators ruled by the set of equations,
\begin{equation}
\left\{
\begin{array}{l}
m \, \frac{d^{2}x_{1}}{dt^{2}}=-\gamma \frac{dx_{1}}{dt}-k^{\prime
}\,x_{1}-k_{1}\left( x_{1}-x_{2}\right) \\ + \eta _{1} \\
\\
m \, \frac{d^{2}x_{i}}{dt^{2}}=-k\,x_{i}-k_{1}\left( 2\,x_{i}-x_{i+1}-x_{i-1}\right) \\
\, \\
m \, \frac{d^{2}x_{N}}{dt^{2}}=-\gamma \frac{dx_{N}}{dt}-k^{\prime
}\,x_{N}-k_{1}\left( x_{N}-x_{N-1}\right) \\ +\eta _{N}%
\end{array}%
\right. ,  \label{mastereq}
\end{equation}
$\left( 2\leq i\leq N-1\right) $, where $\eta $ is Gaussian distributed with,
\begin{equation}
\left\langle \eta _{i}\left( t\right) \,\eta _{j}\left( t^{\prime }\right)
\right\rangle =2\,\gamma \,T_{i}\,\delta _{ij}\,\delta \left( t-t^{\prime
}\right) ,  \label{noises}
\end{equation}
where $T_{i}$ represents the temperature of each reservoir $(i,j=\{1,N\})$. Our analytical approach uses the Stra\-to\-no\-vich interpretation of the noise. The RLL model corresponds to the specific case $k=0$ and $k_{1}=k^{\prime}$, suggesting that the chain is effectively placed as a string pinned only by its extremities onto a base.  More complex potentials can be used~\cite{toda-67}.

Traditionally, the solution for problems of this sort is achieved by assuming a multivariate Fokker-Planck treatment, particularly by making use of the eigenvalue approach~\cite{bergmann-55} or employing Green's function formalism~\cite{dhar-08,kannan-13}. In this manuscript, we handle the problem differently: instead of moving into the probability space and using such methods, we perform our calculations in the Fourier-Laplace space~\cite{dosp-06}. As demonstrated in previous works over the thermo statistics of small systems~\cite{several}, this approach allows obtaining the solid statistical description of a non-equilibrium system in terms of its cumulants --- and the generating function --- skirting the computation of the propagator which might by unreachable for non-linear or non-Gaussian systems.

Explicitly, for the position, we have,
\begin{equation}
\tilde{x}\left( \mathrm{i}\,q+\varepsilon \right) \equiv \int _{0} ^{\infty}  x\left(
t\right) \,\mathrm{e}^{-\left( \mathrm{i}\,q+\varepsilon \right) \,t}\,dt,
\label{position-LF}
\end{equation}
and for the velocity [considering $x _{i} \left( 0 \right) =0$ and $v _{i} \left( 0 \right) =0$ for all $i$ without any loss of generality of our results],
\begin{equation}
\tilde{v}\left( \mathrm{i}\,q+\varepsilon \right) =\left( \mathrm{i}%
\,q+\varepsilon \right) \,\tilde{x}\left( \mathrm{i}\,q+\varepsilon \right) .
\label{velocity-LF}
\end{equation}
As usual, we consider that the system reaches a steady state so that it can be assumed in local equilibrium with the temperature at site $i$, $\mathcal{T}_{i}$, is established by the canonical relation ($k_B =1$),
\begin{equation}
\mathcal{T}_{i} \equiv m\,\left\langle v_{i}^{2}\right\rangle .
\label{temperature}
\end{equation}

The equations of motion~(\ref{mastereq}) can be recast in the form,
\begin{equation}
\mathcal{D}\left( t\right) \,\boldsymbol{x}\left( t\right) =\boldsymbol{\eta
}\left( t\right) ,  \label{operator}
\end{equation}
where $\mathcal{D}\left( t\right) $ is a $N\times N$ operator, \footnote{The form of both operators is made explicit in the Appendix~\ref{contas}.}
$\boldsymbol{x}\left( t\right) $ is the vector of the positions, $\boldsymbol{x}\left( t\right) \equiv \left\{ x_{1}\left( t\right) ,\ldots ,x_{N}\left( t\right) \right\} $ and $\boldsymbol{\eta }\left( t\right) \equiv \left\{ \eta _{1}\left( t\right) ,0,\ldots ,0,\eta _{N}\left( t\right) \right\} $ represents the multivariate stochastic variable describing the fluctuations introduced by the reservoirs. Fourier-Laplace transforming Eq.~(\ref{operator}) we get,
\begin{eqnarray}
\widetilde{\mathcal{D}}\left( \mathrm{i}\,q+\varepsilon \right) \,%
\boldsymbol{\tilde{x}}\left( \mathrm{i}\,q+\varepsilon \right) &=&%
\boldsymbol{\tilde{\eta}}\left( \mathrm{i}\,q+\varepsilon \right) \\
&&  \nonumber \\
\boldsymbol{\tilde{x}}\left( \mathrm{i}\,q+\varepsilon \right) &=&\widetilde{%
\mathcal{A}}\left( \mathrm{i}\,q+\varepsilon \right) \,\boldsymbol{\tilde{%
\eta}}\left( \mathrm{i}\,q+\varepsilon \right) ,  \label{position-operator}
\end{eqnarray}
where $\mathcal{A} \equiv \mathcal{D} ^{-1}$. From Eq.~(\ref{position-operator}) the position of particle $i$ yields,
\begin{equation}
\tilde{x}_{i}\left( \mathrm{i}\,q+\varepsilon \right) = \sum _{j=1,N}
\widetilde{\mathcal{A}}_{ij}\left( \mathrm{i}\,q+\varepsilon \right) \,%
\tilde{\eta}_{j}\left( \mathrm{i}\,q+\varepsilon \right) ,
\end{equation}
and the noise in the reciprocal space, $\tilde{\eta}$, is still Gaussian with
\begin{equation}
\left\langle \tilde{\eta}_{i}\left( \mathrm{i}\,q_{1}+\varepsilon \right) \,%
\tilde{\eta}_{j}\left( \mathrm{i}\,q_{2}+\varepsilon \right) \right\rangle =%
\frac{2\,\gamma \,T_{i}}{\mathrm{i}\,q_{1}+\mathrm{i}\,q_{2}+2\,\varepsilon }%
\delta _{ij}.  \label{noise-LF}
\end{equation}

In this non-equilibrium steady state problem, we can apply the ergodic equivalence between ensemble and time averaging. To benefit from the Fourier-Laplace representation, instead of performing the latter on a stationary stochastic function $f(t)$,
\begin{equation}
\overline{f}\equiv \lim_{\Xi \rightarrow \infty }\frac{1}{\Xi }\int_{0}^{\Xi
}\,f\left( t\right) \,dt,
\end{equation}
we resort to the final value theorem which states that~\cite{vanderpol-50},
\begin{eqnarray}
\overline{f} &=&\lim_{z\rightarrow 0}\,z\int _{0} ^{+ \infty} \exp \left[ -z\,t\right]
\,f\left( t\right) \,dt \\
&&  \nonumber \\
&=&\lim_{z\rightarrow 0,\varepsilon \rightarrow 0}\int _{- \infty} ^{+ \infty} \frac{dq}{2\pi }\frac{%
z}{z-\left( \mathrm{i}\,q+\varepsilon \right) }\,\tilde{f}\left( \mathrm{i}%
\,q+\varepsilon \right)  \label{timeaverage}
\end{eqnarray}
where the Fourier-Laplace representation of $f\left( t\right) $ is used.

The stochastic function $f(t)$ can be taken as the product of stochastic functions as well, e.g., $f(t) \rightarrow v^2(t) = v(t)\times v(t)$. In that case, plugging Eqs. (\ref{temperature}) and (\ref{velocity-LF}) into (\ref{timeaverage}) we obtain,
\begin{eqnarray}
\mathcal{T}_{i} & = & m\lim_{z\rightarrow 0,\varepsilon \rightarrow 0}\int _{- \infty} ^{+ \infty}
\frac{dq_{1}}{2\pi }\frac{dq_{2}}{2\pi }\frac{z}{z-\left( \mathrm{i}\,q_{1}+%
\mathrm{i}\,q_{2}+2\,\varepsilon \right) }  \nonumber \\
& & \nonumber \\
& & \times \left\langle \tilde{v}_{i}\left( \mathrm{i}\,q_{1}+\varepsilon
\right) \,\tilde{v}_{i}\left( \mathrm{i}\,q_{2}+\varepsilon \right)
\right\rangle .  \label{temperature1}
\end{eqnarray}

After some algebra, based on the property given by Eq.~(\ref{noise-LF}), we get the final expression
\begin{equation}
\mathcal{T}_{i} =\frac{m\,\gamma }{\pi } \sum _{j=1,N} T_{j}\int _{- \infty} ^{+ \infty}
q^{2}\, \widetilde{\mathcal{A}}_{ij}\left( \mathrm{i}\,q+\varepsilon \right) \,
\widetilde{\mathcal{A}}_{ij}\left( -\mathrm{i}\,q-\varepsilon \right) \, dq.
\label{temperature2}
\end{equation}

\section{Results}
\label{results}

\subsection*{Answer to Questions 1: On the origin of the cuspidal profile of the RLL model}

\begin{figure}[tbp]
\includegraphics[width=0.49\columnwidth,angle=0]{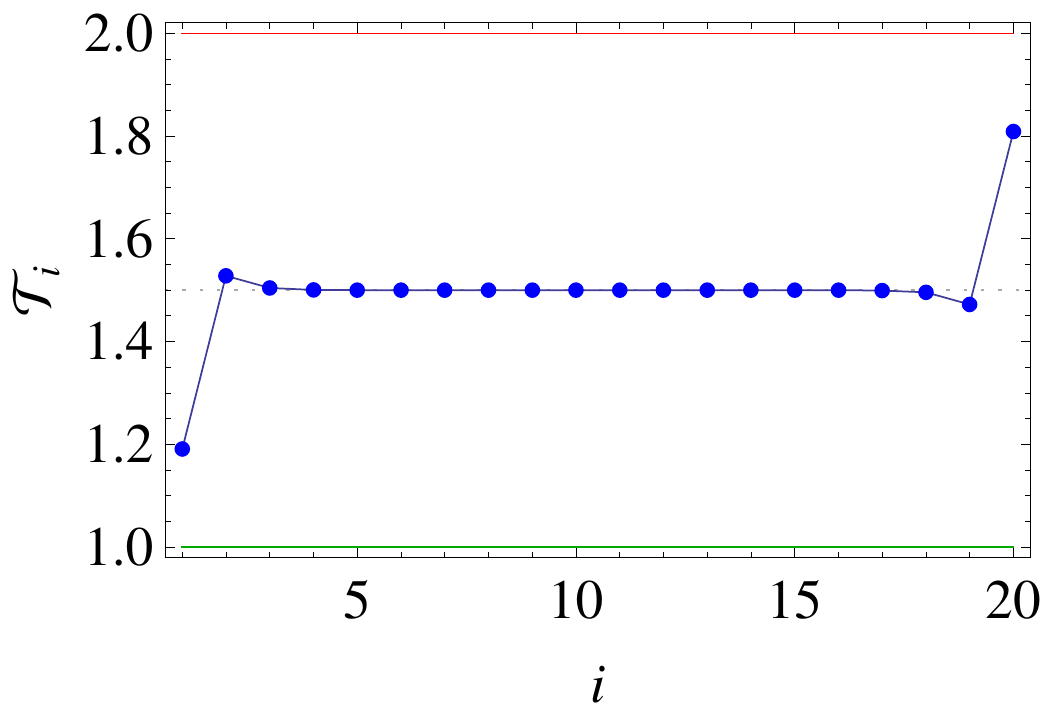} %
\includegraphics[width=0.49\columnwidth,angle=0]{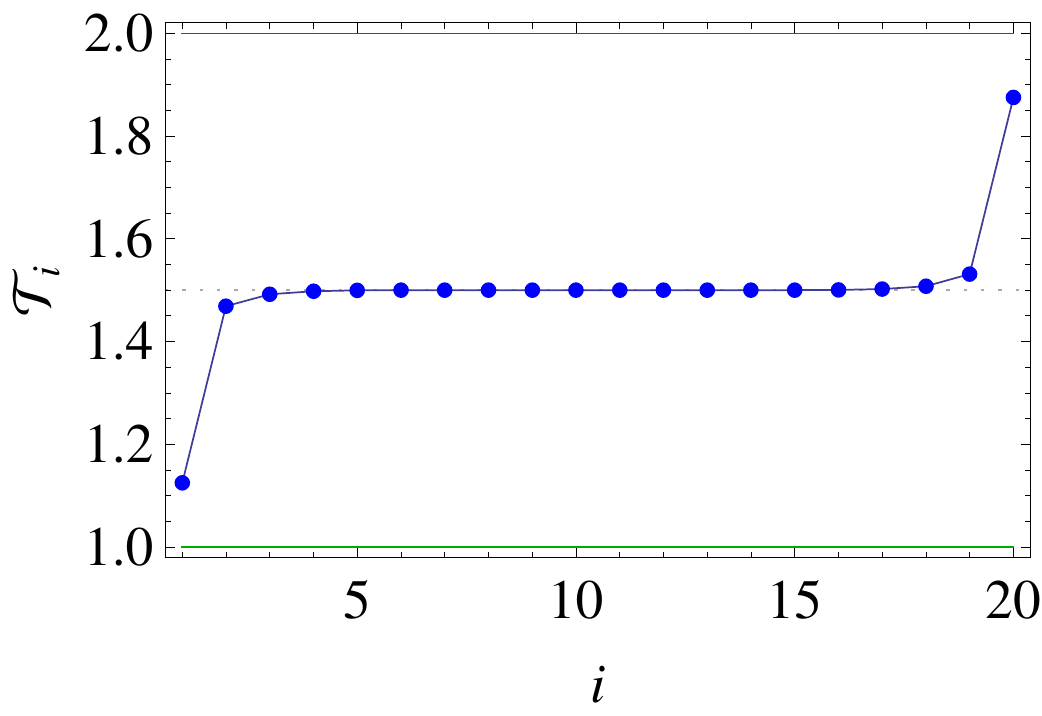} %
\includegraphics[width=0.49\columnwidth,angle=0]{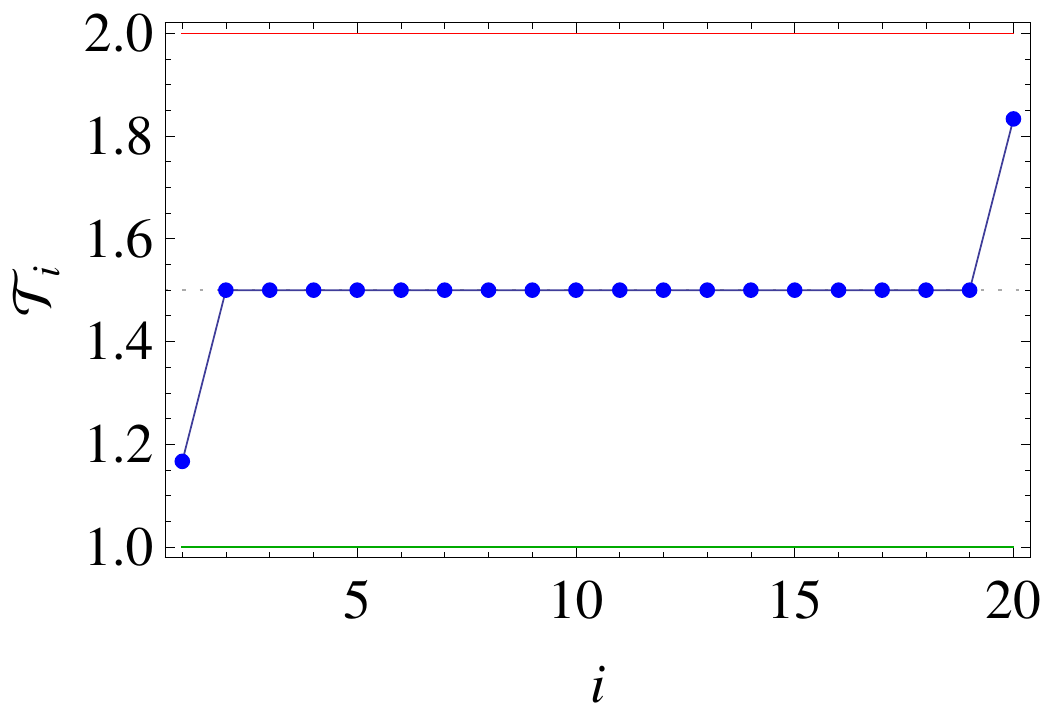}%
\includegraphics[width=0.49\columnwidth,angle=0]{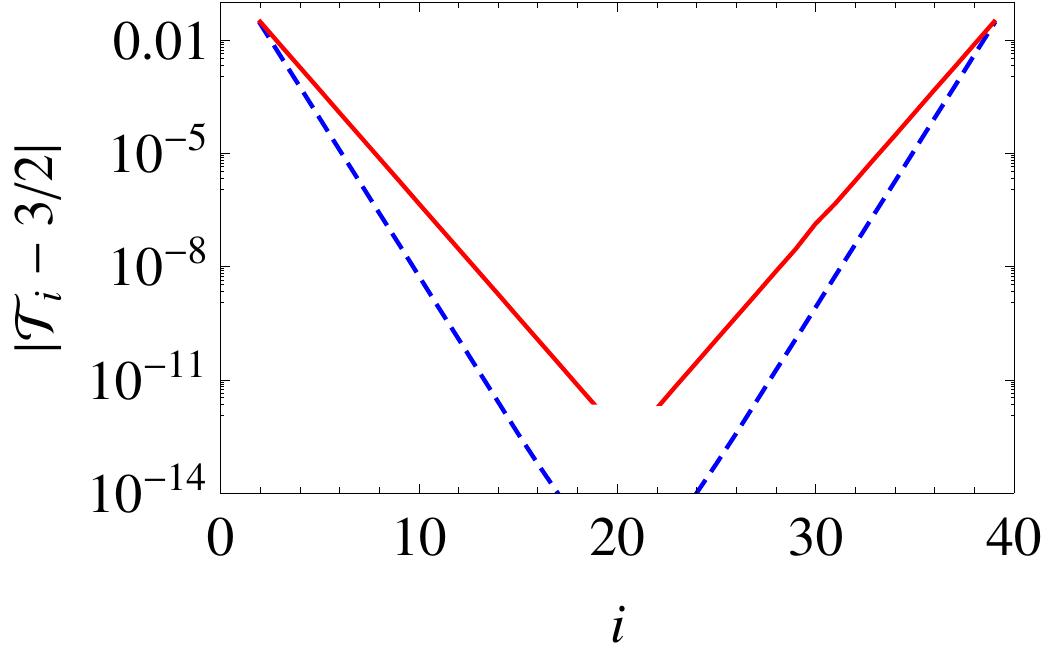} %
\caption{Upper and Lower-left panels: Temperature profile of a chain governed by Eq.~(\protect\ref{mastereq}) with $m=1$, $\gamma = 1$, $k ^{\prime} = k_{1} = 1$ and $k=0$ (UL), $k=1$ (UR) and $k=k_{crit}=1/2$ (LL). The Lower-right panel shows the absolute deviation from mid-temperature versus site position in log-linear scale exposing its exponential dependence. The blue dashed line is for the cuspidal RLL situation $k ^{\prime} = k_{1} = 1$ and $k=0$, whereas the red line represents the normal case with $k=1$.}
\label{temp-prof}
\end{figure}

Matching the conditions of the RLL model~\cite{lebowitz-67} within Eq.~(\ref{mastereq}) we have $k=0$ and $k^{\prime }=k_{1}$ (and $k_3 = 0$). As mentioned, the respective temperature profile presents a cusp (starting) at site $i=2$ and an anti-cusp (ending) at site $i=N-1$ (see left-hand panel in Fig.~\ref{temp-prof}). The absolute value of the departure of the local canonical temperature $\left[\mathcal{T} _i -\left( T _1 + T_N \right) /2\right] $ is known to follow an exponential decay with the distance between the site $i$ and closer end (see the dashed blue line in Fig.~\ref{temp-prof}).

The condition $k=0$ is equivalent to having our chain suspended or laid on a neutral substrate, which either is not always the most typical way of implementing a system or corresponds to a quite simplistic approach~\cite{goodson}. Removing that constraint, we introduce an effective interaction between the bulk and the substrate and by this pinning interaction between the whole chain and the substrate, the (anti)cusps will remain until, for a critical value $k_{\mathrm{crit}}\equiv k_{\mathrm{crit}}(k^{\prime },k_{1})$, the system yields an exact temperature plateau at $\mathcal{T} $ for all particles [except $i=\left( 1,N\right) $] (see lower left panel in Fig.~\ref{temp-prof}). According to our analysis, the critical pinning of the bulk follows the \emph{size-independent} relation,
\begin{equation}
k_{{crit}_1}=\frac{\left( k^{\prime }+k_{1}\right) }{4},\qquad \left( k^{\prime
}=k_{1}\right) .
\label{critical2}
\end{equation}
The size-independence of this relation is understandable when we recall this model is integrable and yields a size-independence (non-decaying with $N$) energy flow.

As we continue increasing the pinning of the bulk, $k$, the plateau is destroyed and there emerges a non-cus\-pi\-dal temperature profile  for which the temperature evolves from the temperature of one reservoir to the other in a monotonically way across the chain (see right-hand panel in Fig.~\ref{temp-prof}). For the smooth temperature profile, we verify the same functional exponential decay of the local canonical temperature with the distance to the respective nearer end of the chain similarly to the RLL case~\cite{lebowitz-67} (see the solid red line Fig.~\ref{temp-prof}). Considering $\Delta k \equiv k - k_{\mathrm{crit}}$, we learn that the characteristic scale of that decay, for $\Delta k < 0$, is larger than the same scale when $\Delta k > 0$.

Thus, the effective origin of the awkward cuspidal temperature profile presented by the Rieder-Lebowitz-Lieb model of heat conduction is related to the absence of interaction between the  substrate and the main physical system, the chain through which the heat flows. 

\begin{figure}[h]
\centering
\includegraphics[width=0.55\columnwidth,angle=0]{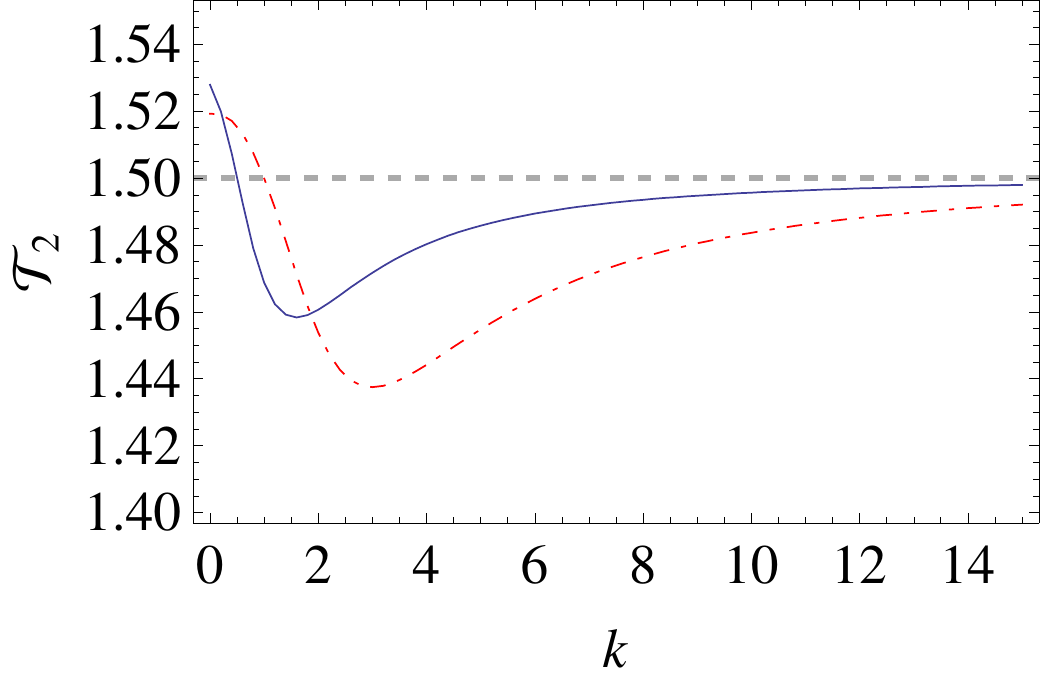}
\caption{Canonical temperature of the second particle of the chain, $%
\mathcal{T} _2$, vs bulk pinning, $k$. with $m=1$, $\protect\gamma = 1$. The
solid blue line is for $k ^{\prime} = k_{1} = 1$ and the red dot-dashed line is
for $k ^{\prime} = k_{1} = 2$. The critical values of $k$ are $k _{crit} = 1/2$ and
$k _{crit} = 1$, respectively, and concur with Eq.~(\ref{critical2}).
In the limit, $k \rightarrow \infty $, the chain turns into an
incompressible wire and the canonical temperature naturally approaches $3/2$.}
\label{t2-rll}
\end{figure}

In Fig.~\ref{t2-rll}, we depict the typical evolution of the temperature of the second particle with respect to $k$ (bulk pinning). In our study, this particle, as well as the $N-1$ particle act as the leading particles of the problem (the first bulk particle); we define them as guiding temperatures (GT).  From that same plot we perceive that $\mathcal{T}_2$ approaches $\mathcal{T}$ from below(above) for $k>(<) k_{crit}$ as given by Eq.~(\ref{critical2}).

\begin{figure}[tbp]
\centering
\includegraphics[width=0.55\columnwidth,angle=0]{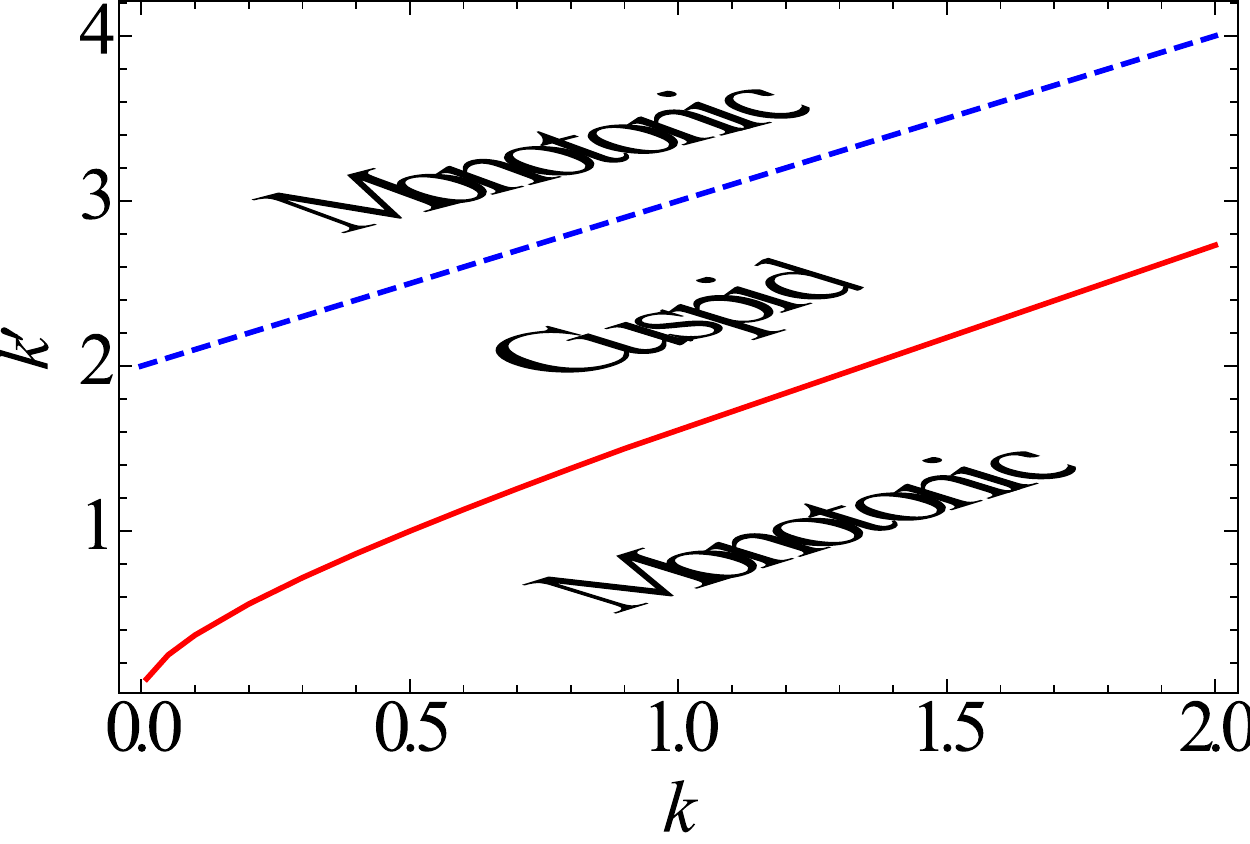} %
\caption{Critical lines defining different behavior for $k_{1}=1$.
The red line indicates the first plateau, $k _{crit_{1}}$, and the blue dashed line is given by
Eq.~(\ref{critical1}). Both lines appear to run parallel for large values of $k$.}
\label{phase-diag}
\end{figure}

Observe that a minimum occurs for ${\cal T}_2$ in Fig.~\ref{t2-rll} (for $k_1= k =1$). The reason is that,  for small values of  $k$, the system will be in the cusp-like local temperature distribution, yielding ${\cal T}_2$ slightly above 1.5 for small $k$, as can be seen in Fig.~\ref{phase-diag}. As one increases $k$, the system switches to the monotonic distribution of temperatures thus bringing ${\cal T}_2$ below 1.50. It keeps decreasing but only up to a point. The reason is that, for extremely large $k$, the system pretty much decouples from the reservoirs (the thermal conductance tends to zero) and the whole system behaves as if in equilibrium, at temperature 1.50. Thus, as $k$ tends to infinity, ${\cal T}_2$ must increase again and reach 1.50. That is the reason why a minimum shows up for the behavior of ${\cal T}_2$

In practice, the RLL model considers the chain and the walls of the reservoirs made of identical materials. If we set loose this condition we allow $k^{\prime} \neq k_{1}$ as well. Accordingly, enlarging the space of parameters, we find two threshold surfaces at which it is possible to obtain an absolutely flat temperature profile. Increasing $k^{\prime }$ from zero, for a given pair of values $\left( k,k_{1}\right)$, we find a first crossover from monotonic to cuspidal behavior of the profile and afterwards a second change to a monotonic profile. For the time being we pass over the first plateau occurring at $k_{crit_{1}}^{\prime }$ --- to which we shall be back shortly --- and mention that the second plateau occurs when the relation
\begin{equation}
k_{crit_{2}}^{\prime }=k + k_{1} + \frac{\gamma^{2}}{m}.  \label{critical1}
\end{equation}
is verified. In Fig.~\ref{phase-diag}, we depict the threshold lines for the case $k_{1}=1$ which represents the general behaviour of the $k_{crit_{1}}^{\prime } \left( k,k_{1}\right)$ surface. 

Despite the objective answer to \emph{Question 1}, the pre\-sent results have shown us that the pinning interaction $k$ between the chain and the substrate  must satisfy $k > k_{\mathrm{crit}_1}$, in order to eliminate the cusps. In other words, the effective origin of the cusps (a characteristic  of the temperature distribution) arises from statistical properties of the dynamics, as will be made more clear in the following. Several factors combine to set the values of $\mathcal{T}_1$ and $\mathcal{T}_{N}$, the most important of them being the heat flux,
\begin{eqnarray}
\mathcal{J} &\equiv& \frac{k_1}{2} \left\langle \left(x_i - x_{i+1} \right)  \left(v_i + v_{i+1} \right) \right\rangle \nonumber\\ &=& - \kappa \, \Delta T \qquad i \in (1,N-1),
\end{eqnarray}
where $\kappa$ represents the \emph{thermal conductance} of the system and $\Delta \, T \equiv T_{H} - T_{C} $, with $T_C$ and $T_H$ being the temperatures of the cold and hot reservoirs respectively.

Intuitively, by increasing the pinning at the edges, we would insulate particles 1 and $N$ from the bulk and make their canonical temperatures $\mathcal{T}_1$ and $\mathcal{T}_N$ (we shall call them the guiding temperatures, GT), respectively, approach the temperature of the reservoirs $T_C$ and $T_H$.  Note that because of the conditions of stationarity
\begin{equation}
\left\langle x_i \, v_i  \right\rangle = \left\langle v_i  \right\rangle = 0.
\end{equation}
Hence, the heat flux is a proxy for the two-point position-velocity correlation function between nearest neighbors, $\mathcal{C}_{x\,v}$.

\subsection*{Answer to Question 2: The statistical behavior of the chain on smoothing its temperature profile}

We first focus on the particles that better indicate the emergence of the plateau, i.e., the bulk leading particles at sites $i=2,N-1$, the GT, and compare the evolution of their canonical temperatures in, $\mathcal{T} ^\ast _{2,N-1}$,\footnote{The asterisk represents that the temperature is defined in $\mathcal{T}$ ($=(T_H+T_C)/2$) units.} to that of steady state heat flux. Without any loss of generality, to compute the steady-state heat flux $\mathcal{J}$, we consider the particle at site $i=1$ and average its power imbalance using Eq.~(\ref{timeaverage}) which equals the $\mathcal{J}$:
\begin{equation}
\mathcal{J}=\lim_{z\rightarrow 0}\,z\int \exp \left[ -z\,t\right] \left\langle \eta _{1}\left( t\right) \,v_{1}\left( t\right) -\gamma
\,v_{1}^{2}\left( t\right) \right\rangle .
\label{imbalance}
\end{equation}
Plugging the relation, $\tilde{v}\left( \mathrm{i}\,q+\varepsilon \right) =\left( \mathrm{i}\,q+\varepsilon \right) \,\tilde{x}\left( \mathrm{i}\,q+\varepsilon \right) $ into Eq.~(\ref{position-operator}), we recast Eq.~(\ref{imbalance}) in reciprocal space,
\begin{equation}
\mathcal{J}={\gamma }^{2} \frac{\Delta \, T}{\pi }
\int \left( \mathrm{i}\,q+\varepsilon \right) ^{2}\,%
\widetilde{\mathcal{A}}_{1N}\left( \mathrm{i}\,q+\varepsilon \right) \,%
\widetilde{\mathcal{A}}_{1N}\left( -\mathrm{i}\,q-\varepsilon \right)
\mathcal{\,}dq.
\label{heatflux}
\end{equation}

\begin{figure}[tbp]
\includegraphics[width=0.49\columnwidth,angle=0]{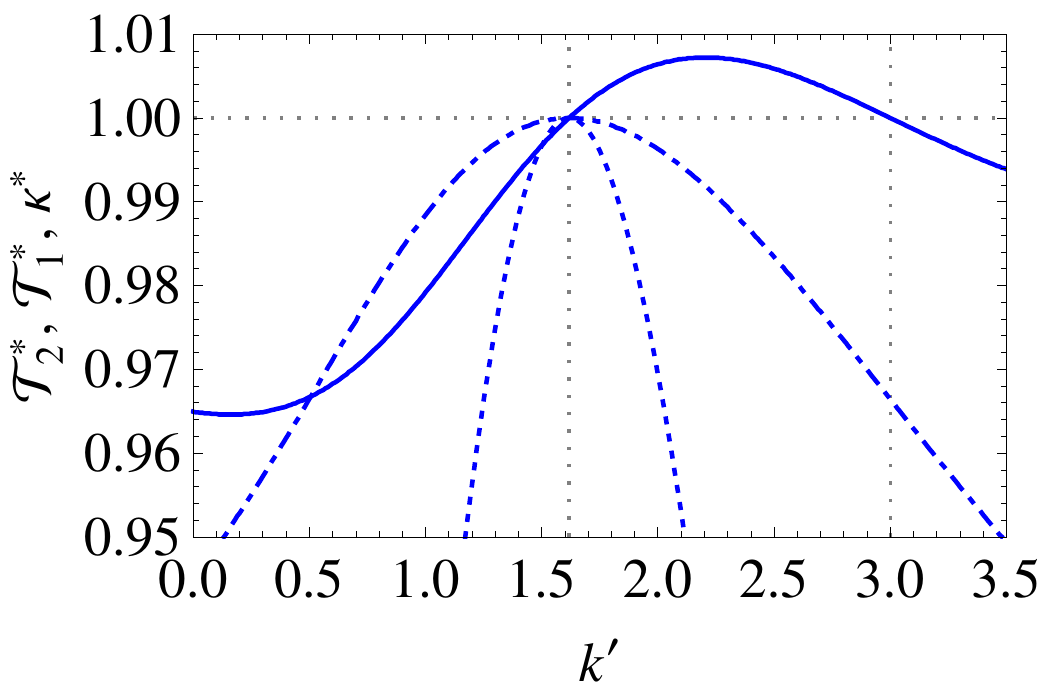}
\includegraphics[width=0.49\columnwidth,angle=0]{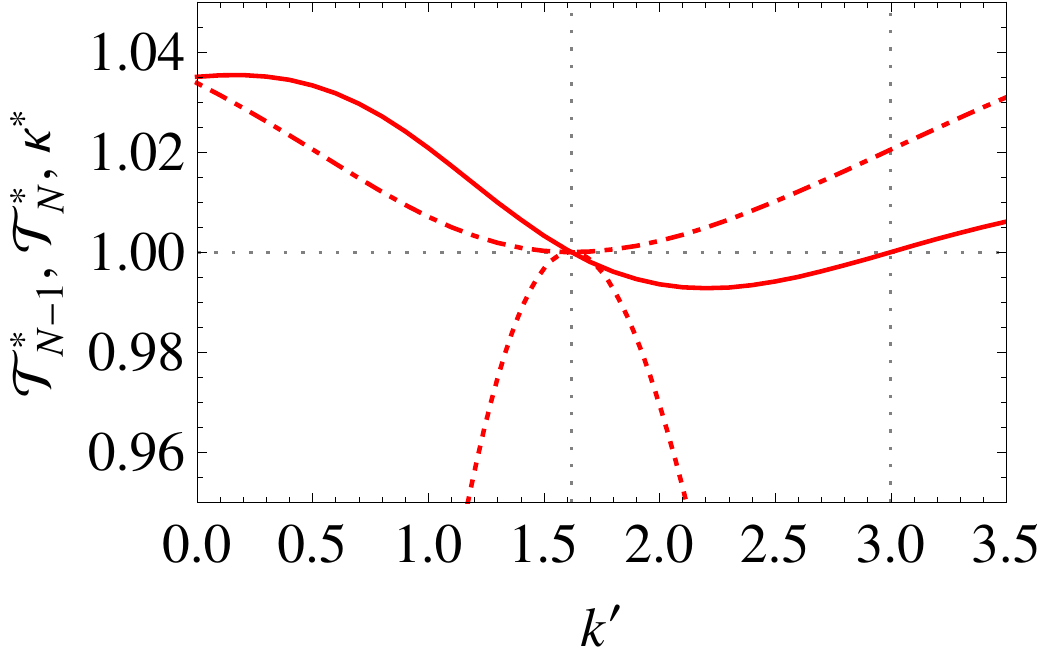}
\caption{Left panel: Normalized values of the canonical temperatures $\mathcal{T}_1$ (dotdashed line), $\mathcal{T}_2$ (solid line) and conductance $\kappa $ (dotted line) vs edge pinning, $k ^{\prime }$, with $m=1$, $\gamma = 1$ and $k = k_{1} = 1$. Both $\mathcal{T}_1$ and $\kappa $ are normalized by the respective maximal values whereas $\mathcal{T}_2$ is normalized by the critical value $\mathcal{T} = 3/2$. The three lines intersect at the critical plateau value $k ^{\prime} _{crit _{1}} =1.618033988749894\ldots$ with value $1$ for all of the three quantities. Right panel: Normalized values of the canonical temperatures $\mathcal{T}_N$ (dotdashed line), $\mathcal{T}_{N-1}$ (solid line)and conductance $\kappa $ (dotted line) vs edge pinning, $k ^{\prime }$, with $m=1$, $\gamma = 1$ and $k = k_{1} = 1$. Temperature $\mathcal{T}_N$ is normalized by its minimal value, $\kappa $ by its maximal values whereas $\mathcal{T}_{N-1}$ is normalized by the critical value $\mathcal{T} = 3/2$. Again, the 
three lines still intersect at the same critical plateau value $k ^{\prime} _{crit_{2}} $ with value $1$ for all of the three quantities. The temperatures $\mathcal{T}_2$ and $\mathcal{T}_{N-1}$ equal the critical value once again for $k ^{\prime} _{crit_{2}} =3 $ as given by Eq.~(\ref{critical1}) and both temperatures approach the value $3/2$ as $k^{\prime } \rightarrow \infty$.}
\label{temp-cond-vs-kl}
\end{figure}

Inspecting the dependence of the flux $\mathcal{J}$ on the sub-space of parameters $\left( k,k^{\prime },k_{1}\right) $ we verify that the heat flux (or the conductance $\kappa = - \mathcal{J}/\Delta T$) reaches its maximal value exactly at the first plateau, see  Fig.~\ref{temp-cond-vs-kl}. Taking into consideration that the heat flux on an infinite  linear chain reads~\cite{dhar-08},
\begin{equation}
\mathcal{J} =\frac{\gamma k_{1}^{2} }{m\, \Theta^{2}}\left(\Pi- \sqrt{\Pi^{2}-\Theta^{2}} \right) \Delta T,
\end{equation}
where,
\begin{equation}
\Theta= 2\, \frac{k_{1}\gamma^{2}}{m}+ 2\,k_{1} \left( k_{1}+k-k^{\prime} \right)  ,\quad
\Pi=  \frac{k \gamma^{2}}{m} + \left(k^{\prime} -k\right)^{2} ,
\end{equation}
we obtain  the equation for the first threshold (solid red line in Fig.~\ref{phase-diag}).
\begin{equation}
k^{\prime}_{crit _{1}} = \frac{k}{2}+ \frac{\sqrt{k^{2}+4 k k_{1}}}{2}.
\end{equation}

\begin{figure}[tbp]
\includegraphics[width=0.49\columnwidth,angle=0]{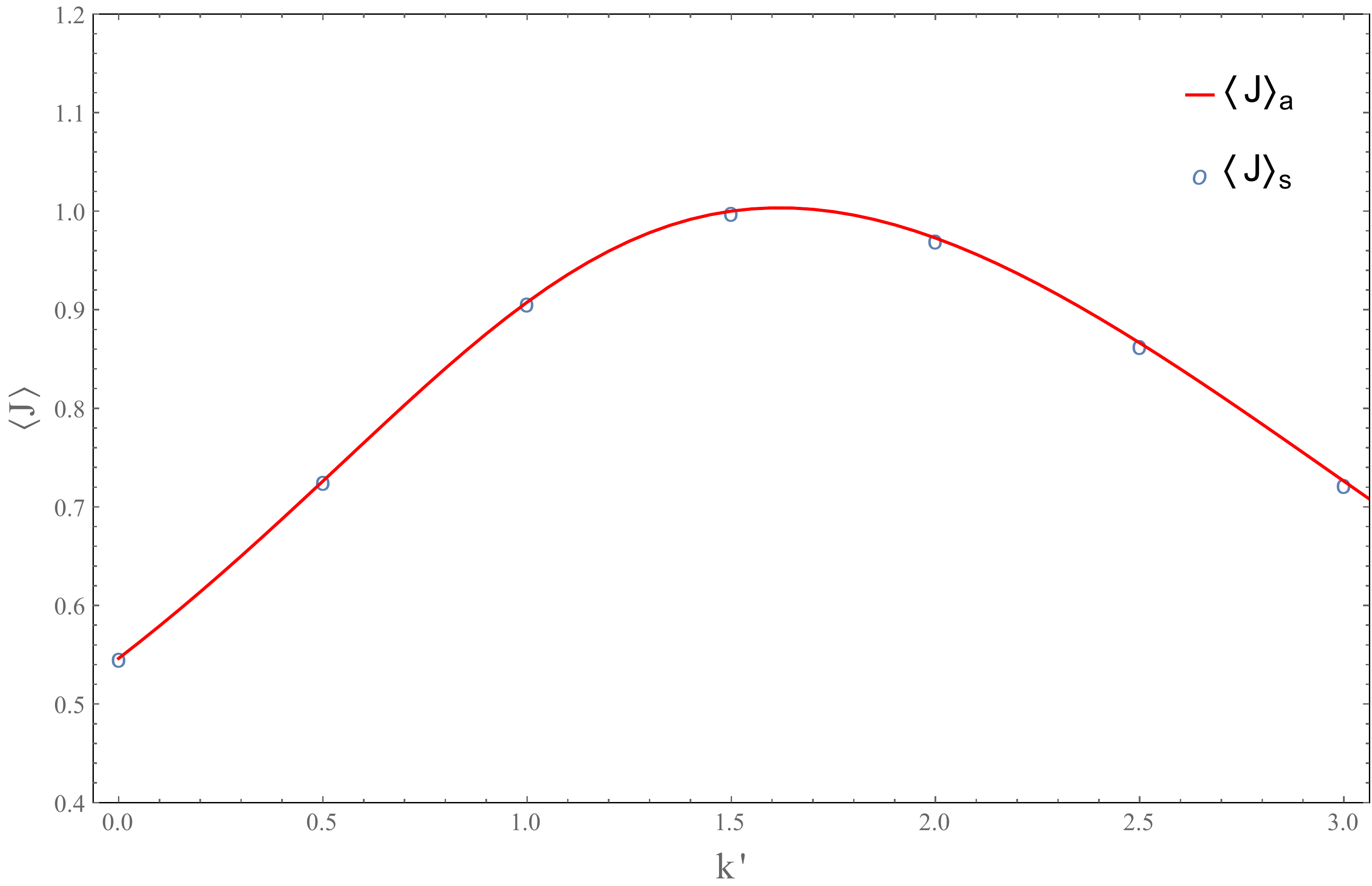} %
\includegraphics[width=0.49\columnwidth,angle=0]{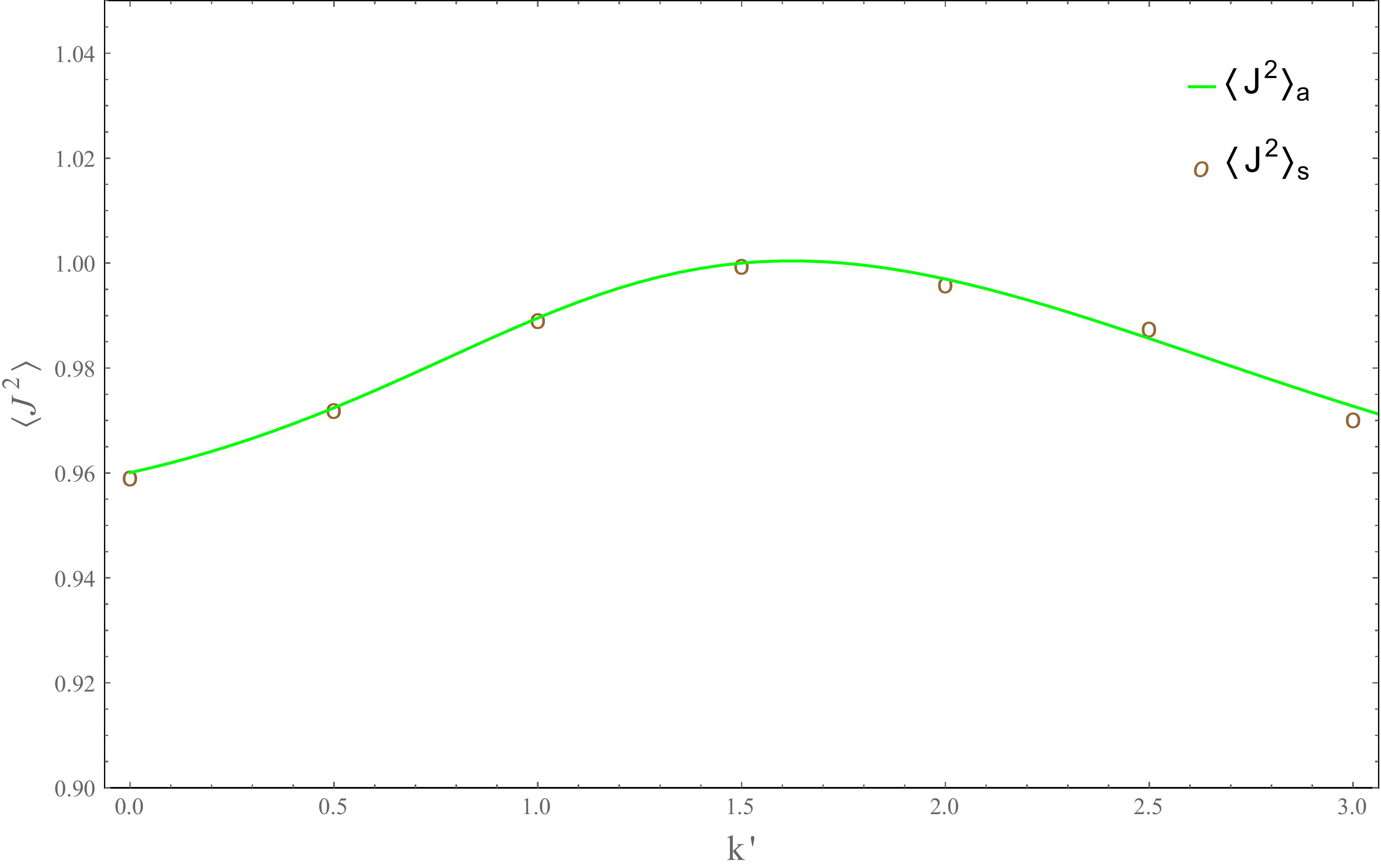} %
\includegraphics[width=0.49\columnwidth,angle=0]{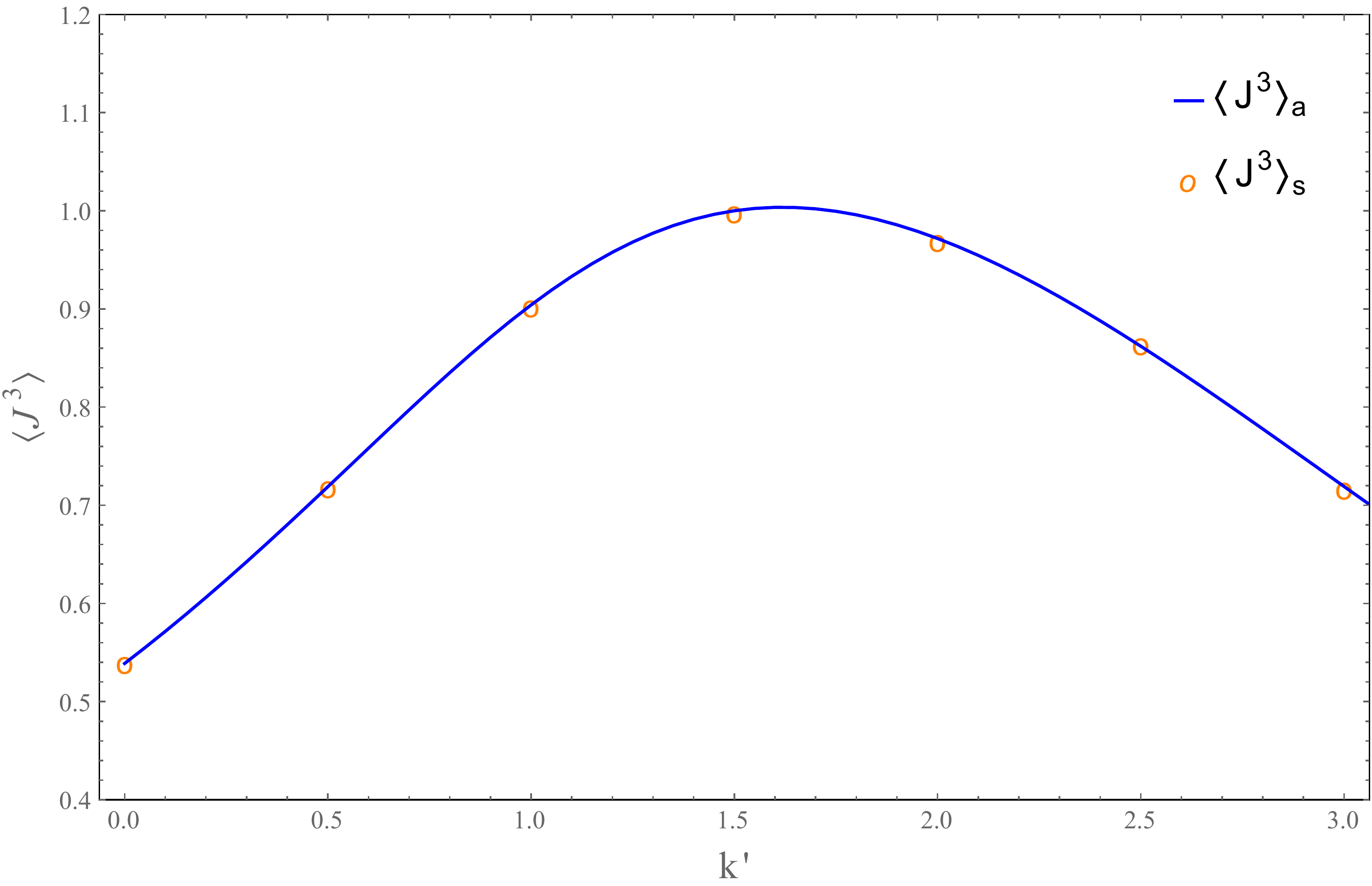}%
\includegraphics[width=0.49\columnwidth,angle=0]{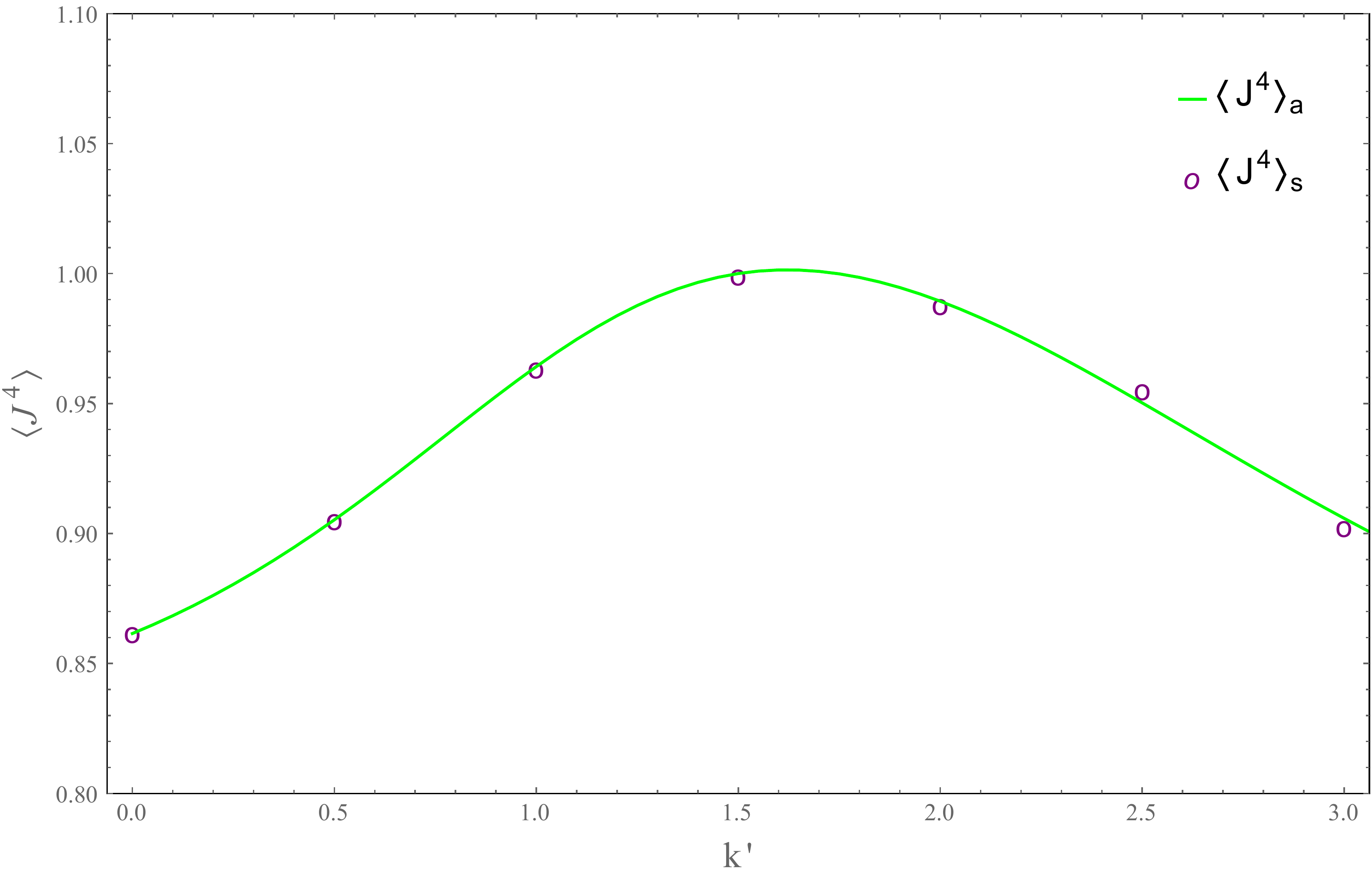}%
\caption{Comparison of numerical and analytical results for the first four moments of heat current  (normalised by the respective maximal value) v edge pinning, $k ^{\prime }$, with $m=1$, $\gamma = 1$ and $k =1$, $ k_{1} = 1$. The upper left and right panels illustrate $\left \langle J \right \rangle$  and $\left \langle J^{2} \right \rangle$, respectively, while the lower left and lower right panels depicts the $\left \langle J^{3} \right \rangle$ and $\left \langle J^{4} \right \rangle$. The solid lines represents analytical results and the circles numerical simulations. For this set of parameters the first crossover occurs at $k ^{\prime} _{crit _{1}} =1.618033988749894\ldots$.}
\label{allcumulants}
\end{figure}

Moreover, we verify that not only the average of the heat flux reaches its maximal value at $k_{crit_{1}}^{\prime }$ but all of its other cumulants as well. That property is depicted in Fig.~\ref{allcumulants} and establishes a clear difference between the two cuspidal-smooth temperature profiles. 
\begin{figure}[tbp]
\centering
\includegraphics[width=0.55\columnwidth,angle=0]{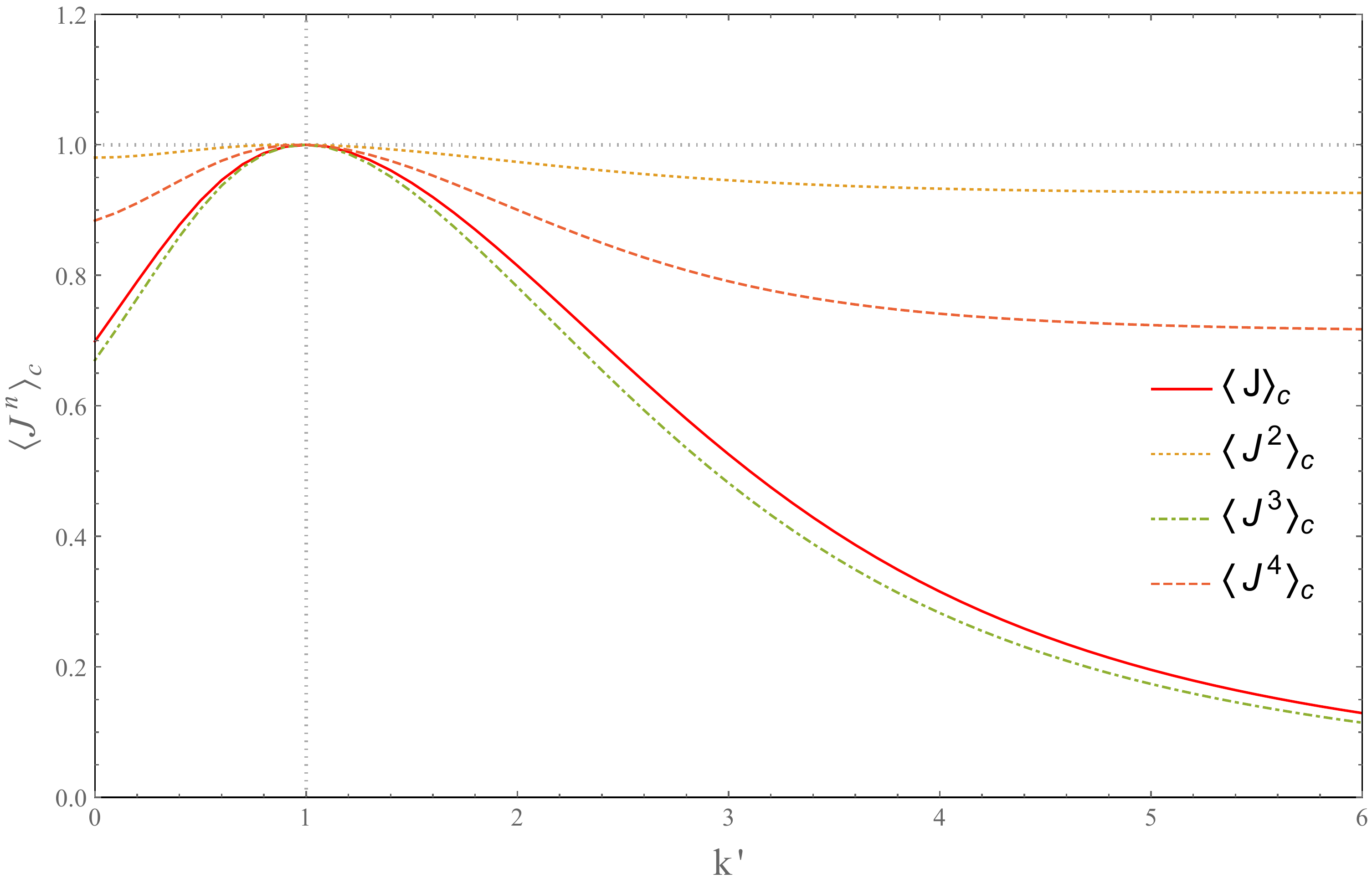}
\caption{The first four order cumulants of the heat flux (normalised by the respective maximal value) v edge pinning, $k ^{\prime }$, with $m=1$, $\gamma = 1$ and $k =\frac{1}{2}$, $ k_{1} = 1$. All the cumulants reach its maximal value $k ^{\prime} _{crit _{1}} =1$. The asymptotic (phonon box) values of $\left\langle J^2 \right\rangle _c$ and  $\left\langle J^4 \right\rangle _c$ are $0.92405279\ldots$ and $0.711240750\ldots$, respectively.}
\label{cumulants-vs-kl}
\end{figure}

The matching of the maxima of the cumulant has, many times, a very particular meaning in condensed matter physics: it points to the emergence of a phase transition. Although we could think of the plateau as the critical state separating the smooth state,
\begin{equation}
\lim _{ N \rightarrow \infty } \frac{1}{N} \sum _{i=2} ^{N/2} \mathcal{T}_i - \mathcal{T}_{N-i+1} < 0
\end{equation}
from the cuspidal state,
\begin{equation}
\lim _{ N \rightarrow \infty } \frac{1}{N} \sum _{i=2} ^{N/2} \mathcal{T}_i - \mathcal{T}_{N-i+1} > 0,
\end{equation}
we rule out such approach because we do not verify another crucial feature that identifies a transition, i.e., the divergence of the correlation length, $\xi $, characterizing the two-point correlation function of the square velocity,\footnote{$v_i ^2$ can be seen as a measure of the instantaneous temperature at site $i$.} $C_{v^2}$ simply does not happen, as can be seen in Appendix~\ref{BBB}.

So, what does actually happen when we go from the cuspidal to the smooth profile and vice versa? We infer that the average population of vibration modes under\-goes a chan\-ge that, it replaces vibration modes more energetically favorable on the colder side with modes more energetic on the hotter side.   The average local energy near the cold source decreases and keeps decreasing as we increase the border pinning $k^{\prime}$. The opposite happens near the hot source. It is the asymmetric behavior of these modes that causes the cusps in the first place. By allowing the system to equilibrate, either by isolating it from the reservoirsor or by taking the external sources to the same temperature, the modes population becomes symmetric and the temperature reaches a true equilibrium plateau.

 As $k^{\prime}$ further increases, the current decreases whereas its fluctuations decrease and the even cumulants will reach eventually their equilibrium values (the ``phonon box'' values which are smaller than the transition point values) while the odd ones vanish at the limit of null heat flux as $k^{\prime}\rightarrow\infty$ (see Fig~\ref{cumulants-vs-kl}).

\section{Concluding Remarks}
\label{remarks}

Inspired by the odd cuspidal temperature profile exhibited by the milestone heat conduction model introduced by Rieder, Lebowitz \& Lieb some 50 years ago~\cite{lebowitz-67} we have shed light on the quantitative relations between the mechanical features, magnitude of the heat flux cumulants and the two-point velocity correlation function for describing the temperature profile of a non-equilibrium chain in contact with two heat reservoirs with different temperatures $T_{C}$ and $T_{H}$ ($T_{C} < T_{H}$). The RLL model is known to bear two independent shortcomings: a ballistic regime of heat transmission with the temperature profile close to a plateau --- which has been intensively studied in the literature --- and a the existence of a cusp (anti-cusp) close to the colder (hotter) reservoir. That profile is reckoned odd since it makes the half of the bulk particles sit closer to the colder reservoir hotter than the other half of the bulk particles that are located next to the hotter reservoirs. We have first shown that the effective reason for the cuspidal profile arises from the lack of interaction between the chain and the substrate. A subsequent analysis has shown that it does not suffice to allow that kind of interaction though; by increasing from $k^\prime=0$ the pinning value at the edges of the chain, we have been able to transition between monotonic and cuspidal behavior and get back former. Therefore, we have shown that in adjusting the cuspidal temperature profile by either increasing or decreasing the pinning in the space of parameters we observe the emergence of a perfect temperature plateau with $\mathcal{T}_{i} = (T_{C} + T_{H}) / 2$ (for all $2 \ge i \ge N-1$).

In further analyzing the thermo statistical properties of the chain, we have found a remarkable property of the local temperature distribution at $ k^{\prime}=k^{\prime}_{criti_{1}}$, where  the cumulants of the heat flux reach extreme values simultaneously. For other values of the parameters choice, with the respective $k^{\prime}_{\mathrm{crit}_{1}}$, and the temperature plateau emerges; particularly, when $k^{\prime} \rightarrow \infty$, the odd cumulants of the heat flux vanish and the even ones reach the values corresponding to a completely isolated system at a given temperature ($=(T_1+T_N)/2$), characteristic of a ``phonon box behaviour''. In that case, we expect the two-point velocity correlation function to zero out due to the symmetry of the Boltzmann-Gibbs distribution.

Early studies on Lyapunov exponents of thermal conductivity models~\cite{mimnagh-97}, namely the Toda model~\cite{toda-67} and the ding-a-ling model~\cite{casati-84} have pointed to a critical-like behavior with finite-size effects. In our case, we tested the critical relations in chains of different sizes and we have found no sensitiveness to the size of the system. Within this context, we have shown that there is no phase transition at all, since the correlation length related to the velocities of the neighboring sites does not diverge at $k^{\prime}_{\mathrm{crit}_{1}}$, presenting a linear behaviour instead. For there is no phase transition at $k^{\prime}=k^{\prime}_{\mathrm{crit}_{1}}$, there must exist another physical reason to justify the shift in the temperature profile. 

The change in temperatures at the lattice extremities tells us about the redistribution of the vibrational modes of the lattice, where it is possible to notice that the half of the chain near the colder reservoir goes from antisymmetric vibrational modes to symmetric vibrational modes, whereas the other half near the hotter reservoir, behaves in the opposite way.

\appendix
\section{Matrix for $\mathcal{D}(s)$ and  $\mathcal{A}(s)$}\label{contas}\label{AAA}
The matrix of dynamics, $\mathcal{D}(s)$, in Laplace space is written as:
\begin{eqnarray}
\mathcal{D}_{i,j}(s) &=& (m s^{2} + \gamma s + k_{1}+k' )\,(\delta_{i,1}\delta_{j,1} +\delta_{i,N}\delta_{j,N}) \nonumber \\
&-& k_{1}\,(\delta_{i,j+1}+\delta_{i,j-1})\,\mbox{  with }  1 \leq i,j \leq N.
\end{eqnarray}
And its inverse is:
\begin{eqnarray*}
\mathcal{A}(s)=
\end{eqnarray*}
\begin{eqnarray*}
= \begin{pmatrix}
A_{11}(s)  &  A_{12}(s)         & A_{13}(s)    & \cdots &  \cdots & A_{1N}(s) \\
A_{21}(s)  & A_{22}(s)        & A_{32}(s)  & \cdots &  \cdots & A_{2N}(s)\\
\vdots     &  \vdots         &  \ddots   & \cdots &  \cdots & \vdots\\
\vdots     &    \vdots       &  \vdots   & \ddots &  \cdots & \vdots\\
\vdots     &      \cdots     & \cdots    & \cdots & \ddots  &  \vdots \\
A_{N1}(s)  &   \cdots        & \cdots    & \cdots &  \cdots &  A_{NN}(s)
\end{pmatrix}
\end{eqnarray*}
The elements $\mathcal{A}(s)_{ij}$ has the structure:
\begin{equation*}
\mathcal{A}(s)_{ij} = \frac{a_{0}+a_{1}s+a_{2}s^{2}+...+a_{2n-3}s^{2n-3}+a_{2n-2}s^{2n-2} }{Det\left[\mathcal{D}(s)\right]}
\end{equation*}
Where $n$ represents the dimension of the chain, and the $a_{k}$ are different constants for each $\mathcal{A}(s)_{ij}$.

Taking the particular case $N=4$ and using the parameters $k'=k=k_{1}=\gamma=m=1$, $\mathcal{D}(s)$ and $\mathcal{A}(s)$ are written,
respectively, as:
\begin{eqnarray*}
\mathcal{D}(s) = \begin{pmatrix}
 s^{2} + s + 2    & -1              &   0         &  0  \\
-1                &  s^{2}+3        &  -1         &  0  \\
0                 &  - 1            &   s^{2}+3   &  -1  \\
0                 &    0            &      -1     &  s^{2} + s + 2
\end{pmatrix}
\end{eqnarray*}
\\
\begin{eqnarray*}
\mathcal{A}(s) = \frac{1}{Det\left[\mathcal{D}(s)\right]}\times
\begin{pmatrix}
\mathcal{A}(s)_{11}   &   \mathcal{A}(s)_{12}               &   \mathcal{A}(s)_{13}                &   \mathcal{A}(s)_{14}   \\
\mathcal{A}(s)_{21}   &   \mathcal{A}(s)_{22}       &    \mathcal{A}(s)_{23}     &  \mathcal{A}(s)_{24}   \\
\mathcal{A}(s)_{31}   &   \mathcal{A}(s)_{32}             &     \mathcal{A}(s)_{33}         &   \mathcal{A}(s)_{34}    \\
\mathcal{A}(s)_{41}   &  \mathcal{A}(s)_{42}        &        \mathcal{A}(s)_{43}             &   \mathcal{A}(s)_{44}
\end{pmatrix}
\end{eqnarray*}
Where $Det\left[\mathcal{D}(s)\right]$ is:
\begin{equation*}
Det\left[\mathcal{D}(s)\right] = s^8+2s^7+11 s^6+16 s^5+40 s^4+38 s^3+54 s^2+26 s+21
\end{equation*}
And for entries we have:
\begin{eqnarray*}
&\mathcal{A}(s)_{11}&=\mathcal{A}(s)_{44}= s^6+s^5+8 s^4+6 s^3+19 s^2+8 s+13 \\
&\mathcal{A}(s)_{12}&=\mathcal{A}(s)_{43}=s^4+ s^3+5 s^2+3 s+5 \\
&\mathcal{A}(s)_{13}&=\mathcal{A}(s)_{42}=2+  s + s^2 \\
&\mathcal{A}(s)_{14}&=\mathcal{A}(s)_{41}=1\\
&\mathcal{A}(s)_{21}&=\mathcal{A}(s)_{34}= s^4+ s^3+5 s^2+3 s+5\\
&\mathcal{A}(s)_{22}&=\mathcal{A}(s)_{33}=s^6+s^5+8 s^4+6 s^3+19 s^2+8 s+13\\
&\mathcal{A}(s)_{23}&=\mathcal{A}(s)_{32}= s^4+2 s^3+5 s^2+4 s+4\\
&\mathcal{A}(s)_{24}&=\mathcal{A}(s)_{31}=2+  s + s^2
\end{eqnarray*}

\section{Correlation length behavior}\label{BBB}

In condensed matter physics, the coincidence in the maxima of the cumulants hints at the existence of a phase transition. Moreover, the emergence of critical behavior in a system is also characterized by the arising of an infinite correlation length, $\xi $, characterizing the two-point correlation function that goes as
\begin{equation}
C_{u}(\delta) \propto \exp \left[ - \delta / \xi \right].
\label{corfun}
\end{equation}
At the plateau, all the bulk particles have the same canonical temperature; hence we have a totally correlated local temperature that could be seen as a sort of ``ordered state of the system''. That said, it is possible to check whether we have a critical-like mechanism by computing the two-point correlation function of the square velocity, $C_{v^2}(\delta)$, for different values of $ k ^{\prime}$ and assess if close to the first threshold we have,
\begin{equation}
\xi \propto \left| \Delta k ^{\prime} \right| ^{-\nu _{\pm}}.
\label{corlen}
\end{equation}
where $\Delta k ^{\prime} = k ^{\prime} - k ^{\prime} _{crit _{1}}$.

\begin{figure}[tbp]
\centering
\includegraphics[width=0.55\columnwidth,angle=0]{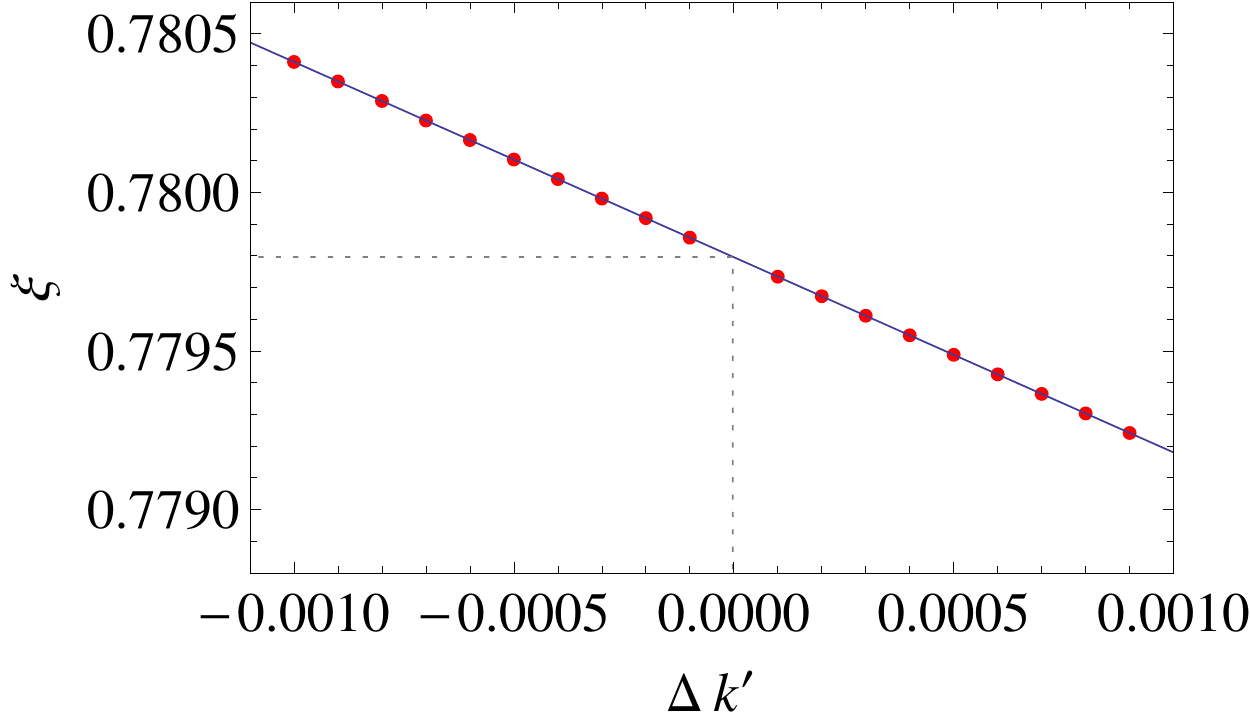} %
\caption{
Correlation length as a function of $\Delta k^{\prime}$. The parameters are the following: $m=k_{1}=\gamma=1$, $k=\frac{1}{2}$, $T_C=1$ and $T_H=2$. The points are obtained from the analytical method and the line corresponds to a linear fit with a slope equal to $-0.61576\pm1.9\times10^{-5}$, ordinate at the origin $\xi ^\ast = 0.779796\pm1.1\times10^{-8}$ and $R^2= 0.99999998$.
}
\label{cor-vv}
\end{figure}

In order to probe Eq.~(\ref{corlen}), we fitted Eq.~(\ref{corfun}) in a log-linear scale for which the slope would be equal to $\xi ^{-1}$. Then, when we pick the correlation length and plot it against $\Delta k^{\prime}$ in a log-log scale we cannot discern a standard critical phenomena power-law; as a matter of fact, we find a quite likely linear dependence ($R^2= 0.99999998$ and $p-\textrm{value} = 10^{-67}$) implying a finite value of $\xi $ for $k ^{\prime} _{crit _{1}}$. The smooth change of $\xi$ lead us to reject the hypothesis of a phase transition scenario.

\begin{acknowledgements}
M.M.C. is grateful to D.K. Foga\c{c}a for valuable discussions.
\end{acknowledgements}

\end{document}